# Supersymmetric mode converters

Matthias Heinrich[1,†,*], Mohammad-Ali Miri[1,*], Simon Stützer[2,*], Ramy El-Ganainy[3], Stefan Nolte[2], Alexander Szameit[2], and Demetrios N. Christodoulides[1]

[1] *CREOL/College of Optics, University of Central Florida, Orlando, FL 32816, USA*

[2] *Institute of Applied Physics, Abbe School of Photonics, Friedrich-Schiller-University,*

*Max-Wien-Platz 1, D-07743 Jena, Germany*

[3] *Department of Physics, Michigan Technological University, Houghton, Michigan 49931, USA*

**Originally developed in the context of quantum field theory, the concept of supersymmetry can be used to systematically design a new class of optical structures. In this work, we demonstrate how key features arising from optical supersymmetry can be exploited to control the flow of light for mode division multiplexing applications. Superpartner configurations are experimentally realized in coupled optical networks, and the corresponding light dynamics in such systems are directly observed. We show that supersymmetry can be judiciously utilized to remove the fundamental mode of a multimode optical structure, while establishing global phase matching conditions for the remaining set of modes. Along these lines, supersymmetry may serve as a promising platform for versatile optical components with desirable properties and functionalities.**

---

[†] *matthias.heinrich@ucf.edu*
[*] These authors contributed equally

The ever-increasing demand for high capacity optical transmission systems[1] has led to remarkable advances in encoding information on a given channel. Wavelength division, polarization and angular momentum multiplexing, multilevel modulation and coherent detection are among the techniques used today in exploiting the various degrees of freedom offered by electromagnetic waves[2,3]. At the same time however, such schemes tend to impose more stringent requirements on the signal-to-noise ratio. Although an increase in the overall transmitted power may improve performance, channel nonlinearities are ultimately expected to be the limiting factor. Mode-division multiplexing (MDM) [4-8] holds great promise in substantially increasing the capacity of optical links[9], while at the same time keeping nonlinearities in check. MDM makes use of the individual modes in an optical waveguide and hence utilizes the available spatial degrees of freedom. One of the outstanding challenges in MDM arrangements is to devise appropriate procedures for selectively populating and extracting specific modes in an integrated fashion. As we will see, Supersymmetry (SUSY) can provide a particularly elegant way to address this issue in a general that is both readily scalable and compatible with existing multiplexing techniques.

The conceptual framework of supersymmetry emerged in the context of quantum field theory as a means to unify the mathematical treatment of bosons and fermions[14-16]. To this end, certain algebraical transformations (see Methods) are employed to construct two different operators that exhibit almost identical eigenvalue spectra. While evidence for supersymmetric behavior in any physical setting, whether natural or artificial, has so far remained elusive, its fundamental ideas can in principle be adopted in other areas of physics[17]. In its optical manifestation, supersymmetry can potentially establish close relationships between seemingly different dielectric structures[10]. For example, two refractive index profiles interrelated via supersymmetric transformations share identical scattering characteristics[11] and therefore can become virtually indistinguishable to an external observer, even in the presence of losses[12,13]. In the context of guided wave optics, supersymmetric partner waveguides are characterized by perfect global phase matching conditions: With the exception of the fundamental mode, each guided mode of the original multimode waveguide has a counterpart in the partner arrangement with exactly the same propagation constant, or effective index (see Figs. 1(a,b)). The refractive index distribution of the superpartner waveguide can be found through a systematic deformation of the original structure (see Supplementary Fig. 1). Yet, to this date, no experimental observation of supersymmetric optical behavior has been reported.

In this article, we demonstrate that the perfect phase matching conditions afforded by supersymmetry can be effectively utilized for multiplexing/demultiplexing the modal content of highly multimoded systems. This is accomplished by judiciously introducing superpartner structures capable of directing different modes from, or to, specific output ports (Fig. 1(c)). A hierarchical ladder arrangement of such superpartners, that can simultaneously interrogate the entire modal content of a system, is experimentally realized. In this respect, mode division multiplexing based on optical supersymmetry (SUSY-MDM) can be performed, in a scalable fashion, simultaneously for a great number of modes, without the need for any additional beam shaping components.

**Results**

*Supersymmetric optical structures*

In what follows, we demonstrate supersymmetric mode conversion in multicore photonic lattice systems. To implement these structures, we employ the femtosecond laser writing technique to inscribe arrays with appropriate index profiles in fused silica (see Methods and Supplementary Fig. 3). From an experimental perspective, the physical platform presented here has a unique advantage over other realizations: The evolution dynamics in such lattices can be observed by means of waveguide fluorescence microscopy[18], hence allowing one to directly evaluate their response. In principle however, our results are general and other fabrication approaches can also be pursued. In all possible settings, light dynamics are dictated by the refractive index distribution. For the photonic lattices employed in this study, coupled-mode theory provides an effective approach in describing light evolution[19]. In this context, light propagation can be discretized[20-22], and as a result, the corresponding state vector **A** obeys the following evolution equation along the longitudinal coordinate $z$:

$$-i\frac{d}{dz}\mathbf{A} = \mathcal{H}\mathbf{A} . \qquad (1)$$

Here, $\mathbf{A} = (a_1, \ldots, a_N)^T$, where $a_k$ describes the complex modal field amplitude in the $k^{\text{th}}$ channel, $N$ is the number of lattice sites involved, and the $N \times N$ matrix $\mathcal{H}$ is the Hamiltonian of the system whose elements are given by $\mathcal{H}_{m,n} = (\delta_{m-1,n} + \delta_{m+1,n})C_n + \delta_{m,n}\beta_n$. In the latter expression, $\beta_n$ denotes the propagation constant of channel $n$, and $C_n$ represents the coupling strength between adjacent lattice sites. Note that our fabrication method provides full control over these elements[23]. The eigenvalue problem $\mathcal{H}\mathbf{A} = \lambda\mathbf{A}$ associated with Eq. (1) can in turn be used to construct a superpartner lattice (see Methods and Supplementary Fig. 2). As an example, Fig. 2(a) shows the refractive index profile of a fundamental lattice involving six identical sites, while Fig. 2(b) depicts its discrete representation. Similarly, Figs. 2(c,d) illustrate the corresponding superpartner index landscape and the associated array, consisting of five sites as a result of unbroken supersymmetry. Note that the superpartner shares a common set of propagation constants (eigenvalues) with the original structure, with the exception of that of the fundamental mode (Fig. 2(e)). To factorize the discrete operators involved, we use Cholesky's method as well as the so-called QR decomposition – the discrete counterparts to the continuous supersymmetric transformations.

To elucidate the fundamental principle behind optical supersymmetry, let us first consider the simplest possible case where a structure supports only two bound modes (Fig. 2(f)). In general, light injected into this waveguide will populate both of these states, and as a result the ensuing interference leads to a periodic propagation pattern. An experimental observation of such a bimodal beating is shown in Fig. 2(h). By applying a supersymmetric transformation, one can then establish a partner refractive index profile. For this particular case it turns out that the resulting structure is single-moded (i.e. supports only the ground state), and is exactly phase-matched to the second mode of the original waveguide (Fig. 2(g)). Naturally, light injected into the superpartner here displays stationary evolution, as shown in Fig. 2(i).

*The hierarchical SUSY ladder*

The fact that supersymmetry can establish global phase matching conditions among the bound states of two different partner potentials brings about the possibility of successively isolating and extracting these modes – a necessary attribute for MDM schemes. This property can in turn be exploited to control light transport in a hierarchical "ladder" of iteratively generated superpartners. In other words, the number of modes supported by each "step" in this ladder is sequentially reduced, until only a single bound state (or waveguide channel) remains. Such a ladder is schematically depicted in Fig. 3(a). In this example, starting from a fundamental multimode structure supporting six states, a progression of five partner structures is obtained, each of which is supersymmetric with respect to its immediate predecessor. As a result, the sets of eigenvalues corresponding to the individual step of the ladder are perfectly aligned. For instance, the third mode of the fundamental structure can exchange energy with its counterparts in the second and third step, while it is prohibited from interacting with the others. Notably, coupling between the superpartners breaks the degeneracy around these eigenvalues, giving rise to multiplets (or bands) of collective ladder states as illustrated in Fig. 3(a). Along these lines, SUSY ladders can in principle allow one to simultaneously multiplex a massive number of modes with a single operation. Note that supersymmetric phase-matching is robust even in the presence of a partner waveguide. Numerical simulations (see Supplementary Discussion and Supplementary Fig. 4) of a continuous six-moded waveguide with a step-like index profile and its superpartner show that the ensuing crosstalk between non-phase-matched modes remains below 20 dB, in spite of the fact that the index profiles partially overlap. With a more generous spacing, or by excluding the mode pair closest to the cutoff, conversion fidelities of 35 dB are readily achievable and modes can be effectively transformed across the entire ladder (see Supplementary Fig. 5).

*Mode conversion and isolation*

In order to observe this behavior, we implemented a SUSY ladder in fused silica glass. In our arrangement, the fundamental structure supports six modes, similar to the schematic representation in Fig. 3(a). The coupling between successive superpartners is achieved by placing them in close proximity to one another, allowing for evanescent energy transfer. The fundamental state in each respective step is excited by launching a Gaussian beam perpendicularly to the input facet. Figures 3(b-g) show the experimentally observed propagation dynamics arising from such excitations. Indeed, light injected in the ground state of the fundamental array remains localized, and is completely isolated from the rest of the ladder (Fig. 3(b)). In contrast, wave packets originating from the ground state of any higher layer can freely traverse the ladder and are directed towards the fundamental partner.

On the other hand, higher order excitations of the fundamental system (e.g. in the $k^{\text{th}}$ state, $k > 1$) can be transported across the ladder, up to the corresponding step $\ell = k$. This is confirmed by experimental results, shown in Fig. 4(a-c), where the input beam was appropriately tilted so as to selectively populate mixtures of the three lowest states. Note that, in all cases, the output patterns clearly reflect the mode transformation that takes place in the SUSY ladder. The conversion becomes apparent from the node-free distribution in the respective highest accessible steps. Our results indicate that such specifically designed supersymmetric arrangements could be useful for efficiently probing, manipulating and interrogating the modal content of a given input field distribution. This is enabled by the fact that in such a setup the modes can be spatially separated via global phase matching conditions as afforded by SUSY

– even in highly multimoded systems. In essence, the proposed ladder arrangement incorporates the functionalities of a set of mode converters and successive beam combiners into a single multiplexing component (see Fig.1(c)). This operation is fully reversible, i.e. the same element can be employed to demultiplex the superposition of modes after transmission. We would also like to emphasize that, even though the experimental results presented here are of qualitative nature and were obtained in discrete one-dimensional settings, the fundamental principle of SUSY-MDM is equally applicable to continuous arrangements and can even be extended to optical fibers, where whole subsets of modes may be selectively manipulated according to their specific optical angular momenta[10]. Along these lines, detailed simulations of supersymmetric mode conversion in continuous refractive index landscapes, and its robustness with respect to dispersion, are provided in the Supplementary Discussion (see also Supplementary Figs. 6,7). Moreover, in discrete configurations, one can directly access any subset of modes through higher order superpartners, by means of discrete factorization methods.

It is worth noting that other avenues for manipulating modes have been recently proposed. These include for example "photonic lanterns" that can efficiently transfer light from highly multimoded waveguides to a large number of identical single-mode channels[24], as well as individually phase-matched single-mode cores[25] and adiabatic transitions between dissimilar fiber geometries[26]. What sets our strategy apart from these approaches is the fact that it is readily scalable to large numbers of modes, which can be simultaneously phase-matched in a compact structure.

**Discussion**

In summary, we have experimentally investigated for the first time light transport in supersymmetric optical structures. Apart from providing a physical setting where the unusual ramifications of supersymmetry can be directly explored and investigated, optics offers the opportunity to exploit some of its intriguing features. Our results demonstrate that superpartner potentials can be effectively employed to judiciously manipulate the modal content of an optical field. In particular, hierarchical ladders of supersymmetric partners provide a versatile method for mode discrimination and mode division multiplexing across technological platforms. The inherent scalability of SUSY-MDM becomes even more apparent in highly multimoded environments. This is due to the fact that mode transformation is naturally carried out at the physical layer, thus overcoming the need for separate mode converters or beam combiners (as schematically shown in in Fig.1(c)). Along these same lines, additional degrees of freedom[12] could be utilized in addressing other design goals. SUSY phase-matching can be employed to facilitate high-fidelity mode conversion over a broad spectral range, e.g. throughout the telecommunication-relevant C-band (see Supplementary Fig. 6), and is therefore fully compatible with established wavelength division multiplexing schemes. Similar strategies may also provide a new avenue in realizing orthogonal mode converters, and self-aligning universal beam couplers[27,28]. Finally, notions from supersymmetry can in principle be utilized to synthetize artificial optical structures that exhibit properties not found in nature, thus supplementing already existing approaches in optical metamaterials based on transformation optics[29-32].

**Methods**

*Mathematical framework*

Supersymmetry endows two otherwise unrelated operators, $\mathcal{O}^{(1)}$ and $\mathcal{O}^{(2)}$, with almost identical eigenvalue spectra[14,17,33]. In general, such a relationship exists between these two entities, provided that $\mathcal{O}^{(1)}$ can be decomposed in terms of another operator, $\mathcal{A}$, and its Hermitian adjoint, $\mathcal{A}^\dagger$, in the following manner: $\mathcal{O}^{(1)} = \mathcal{A}^\dagger \mathcal{A}$. In this respect, the superpartner $\mathcal{O}^{(2)}$ can be introduced via $\mathcal{O}^{(2)} = \mathcal{A}\mathcal{A}^\dagger$, and as a result the eigenvalue problems $\mathcal{O}^{(1,2)} Y^{(1,2)} = \Lambda^{(1,2)} Y^{(1,2)}$ share a common set of eigenvalues $\Lambda_j^{(1)} = \Lambda_j^{(2)}$. In all cases, the eigenfunctions $Y_j^{(1,2)}$ corresponding to these spectra are linked by the intervening SUSY operators $\mathcal{A}, \mathcal{A}^\dagger$. Importantly, unbroken SUSY also demands that the ground state of the first operator must be annihilated by $\mathcal{A} Y_1^{(1)} = 0$. Indeed, what sets such a pair of superpartners apart from other systems obeying more conventional symmetries is the fact that the ground state of $\mathcal{O}^{(1)}$ is exempt from this interrelation, and is therefore absent from the spectrum of $\mathcal{O}^{(2)}$ (see Figs. 1(a,b)).

*Factorization of the continuous Hamiltonian*

Under paraxial conditions, the evolution of light in one-dimensional settings is governed by the wave equation

$$\left(i\frac{\partial}{\partial Z} + \frac{\partial^2}{\partial X^2} + V(X)\right)\psi = 0, \qquad (2)$$

where $X = x/x_0$ and $Z = z/2k_0 n_0 x_0^2$ are the transverse and longitudinal coordinates normalized with respect to an arbitrary transverse length scale $x_0$ and the vacuum wave number $k_0$. The optical potential $V = 2k_0^2 n_0 \cdot \Delta n(x)$ is then determined by the refractive index profile $n(x) = n_0 + \Delta n(x)$. A SUSY partner potential can be found by factorizing the operator $H = \partial^2/\partial X^2 + V$ in the eigenvalue problem $H\psi = \mu\psi$ by means of the superpotential method[10] (see Supplementary Fig. 1). Along these lines, the superpotential $W = -\partial/\partial X (\ln \psi_1(X))$ is obtained as logarithmic derivative of the fundamental mode $\psi_1^{(1)}(X)$ of the original structure $\Delta n^{(1)}$. The corresponding SUSY partner index distribution is then given by $\Delta n^{(2)} = \Delta n^{(1)} - 2\, \partial W/\partial X$.

As is shown in Supplementary Fig. 2(a,b) for the case of six identical single-mode channels, the superpartner of a photonic lattice in turn represents a lattice with one less waveguide, although its structure may feature index depressions and in general can no longer be decomposed into identical unit cells.

*Factorization of the discrete Hamiltonian*

In the tight binding approximation, the evolution of guided light in a photonic lattice is described by Eq. (1) of the main manuscript. The respective eigenvalue problem can then be written in the form $\mathcal{H}a = \lambda a$, where the discrete Hamiltonian $\mathcal{H}$ is a Hermitian operator composed of the propagation constants $\beta_n$ and the coupling coefficients $C_{n,n+1} = C_{n,n-1} \equiv C_n$. Cholesky's algorithm (see e.g. Ref. [34]) allows for the decomposition $\mathcal{H}^{(1)} = \mathcal{H} - \lambda_1 = \mathcal{A}^\dagger \mathcal{A}$ of positive-definite operators into a Hermitian adjoint pair $\mathcal{A}^\dagger, \mathcal{A}$. The partner Hamiltonian $\mathcal{H}^{(2)} = \mathcal{A}\mathcal{A}^\dagger$ again formally represents a photonic lattice with $N$ waveguides and an equal number of bound modes. Nevertheless, SUSY is unbroken in the sense that the $N^{\text{th}}$ waveguide is completely detached ($C^{(2)}_{N-1,N} = C^{(2)}_{N,N-1} = 0$). It can be discarded without influence on the spectrum or mode shapes of the remaining system of $N-1$ channels, thereby removing the counterpart of the original lattice's fundamental mode from the superpartner's spectrum. Higher order partner Hamiltonians can be synthesized by iteratively eliminating the modes of a given structure: $\mathcal{H}^{(\ell+1)} - \lambda_1^{(\ell)} = \mathcal{A}^{(\ell)}\left(\mathcal{A}^{(\ell)}\right)^\dagger$ where $\mathcal{H}^{(\ell)} - \lambda_1^{(\ell)} = \left(\mathcal{A}^{(\ell)}\right)^\dagger \mathcal{A}^{(\ell)}$. The bands comprised of the collective states in a weakly coupled sequence of such systems exclusively span the layers that support the corresponding eigenvalue (see Fig. 3(a)).

A generalized manipulation of the eigenvalue spectrum can be achieved by means of the so-called QR factorization, which expresses the Hamiltonian as product of an orthogonal matrix $\mathcal{Q}$ and an upper triangular matrix $\mathcal{R}$ (see e.g. Ref. [34]). This asymmetric approach allows for the direct removal of any eigenvalue $\lambda_k$ by factorizing $\mathcal{H}^{(1)} = \mathcal{H} - \lambda_k = \mathcal{Q}\mathcal{R}$ to obtain $\mathcal{H}^{(2)} = \mathcal{R}\mathcal{Q}$. Note that in continuous 1D settings, complex-valued potentials are required to address states other than the fundamental mode[12,13]. In contrast, the QR formalism allows one to accomplish this task without resorting to non-Hermitian configurations involving the interplay between gain and loss.

*Experimental techniques and parameters*

Pulses from a Titanium:Sapphire amplifier system (Coherent Inc. Mira/RegA, wavelength 800 nm, pulse length 200 fs, repetition rate 100 kHz) were focused through a microscope objective ($25\times$, $NA = 0.35$) to inscribe[35] the waveguide arrangements used in our experiments (see Supplementary Fig. 3(a)). The 100 mm long fused silica sample (Corning Inc.) was translated by means of a high-precision positioning system (Aerotech Inc.). In order to cover a wide range of effective refractive indices and nearest-neighbor couplings, we exploited the characteristic evolution behavior of light in detuned directional couplers[36] (Supplementary Fig. 3(b)) to calibrate the dependence of detuning and coupling coefficient on the inscription parameters writing velocity and waveguide separation[37]. In a directional coupler, i.e. a pair of coupled waveguides, light undergoes sinusoidal oscillations between the two channels. In the perfectly tuned case, a complete transfer is observed after the coupling length $L_C = \pi/2C$. As the detuning $\Delta\beta$ between the channels increases, the exchanged fraction of light decreases, resulting in a decreased intensity beating contrast $K \equiv \max(I_1) - \min(I_1)/\max(I_1)$. At the same time, the oscillation becomes more rapid and the beating period $L_B$ decreases according to

$$\frac{\Delta\beta}{C} = \sqrt{\frac{1}{K} - 1} \qquad \text{and} \qquad \frac{L_B}{L_C} = \sqrt{\frac{1}{(\Delta\beta/C)^2 + 1}} \ . \qquad (3)$$

The desired parameters $\Delta\beta$ and $C$ can therefore be calculated directly from the observed quantities $L_B$ and $K$ according to

$$C = \frac{\pi}{2L_B} \cdot \sqrt{K} \qquad \text{and} \qquad \Delta\beta = \frac{\pi}{2L_B} \cdot \sqrt{1 - K} \ . \qquad (4)$$

The calibration graphs thus obtained are shown in (Supplementary Fig. 3(c)). For our experiments, we chose a value of $C_0 = 0.65\ cm^{-1}$, corresponding to a spacing of 15.5 µm, to implement the SUSY ladder arrangement. A writing velocity of 100 mm/min was chosen as baseline corresponding to zero detuning. This allowed us to design an experimental configuration (Supplementary Fig. 3(c)) to observe the desired propagation dynamics over the sample length of 100 mm.

In addition to recording the output distribution at the sample's end facet (see inset in Supplementary Fig. 3(a)), we employed waveguide fluorescence microscopy[18] to directly observe the propagation of light through the structures. At a probe wavelength of 633 nm (Helium:Neon laser), this technique makes use of certain color centers formed during the inscription process to linearly convert a small fraction of the guided light into isotropic fluorescence. The images thus obtained were post-processed to extract the intensities in the individual channels, a necessary step to facilitate a quantitative comparison between SUSY partner lattices despite their different waveguide positions. Moreover, plots of the mode intensities improve the visibility of the propagation dynamics (see Supplementary Fig. 3(e)).

**Figures:**

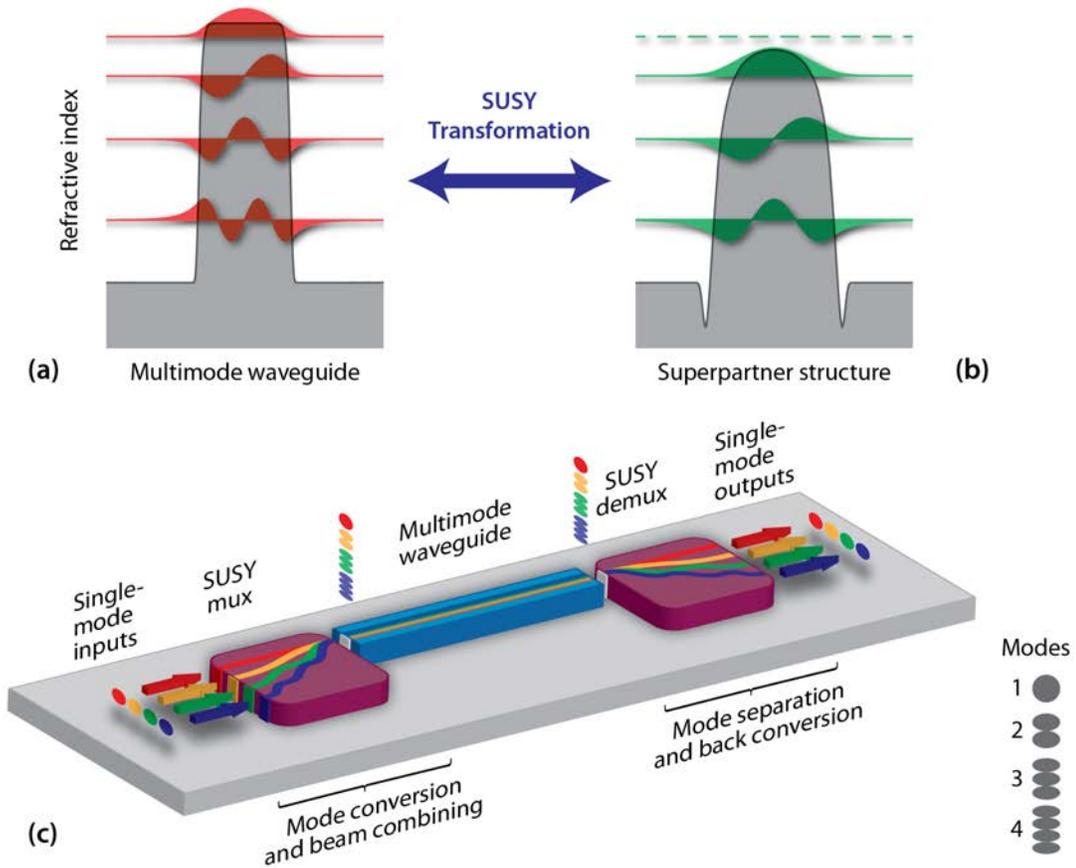

Fig. 1: Supersymmetric optical structures and their application for mode division multiplexing. (a) A multimode optical waveguide supporting four guided states. The three modes of the superpartner (b) are perfectly phase-matched to the higher modes of the original structure. The profiles of the SUSY mode pairs in those two waveguides are related via SUSY transformations. Note that the fundamental mode in (a) has no counterpart in the superpartner structure. For more details, see Methods and Supplementary Fig. 1. (c) Supersymmetric mode division multiplexing. A hierarchical sequence of multiple superpartner structures incorporates the functionalities of successive mode converters and beam combiners into a single multiplexing/demultiplexing component (SUSY mux/demux). All input- and output-ports of this schematic arrangement operate in their respective fundamental modes. The individual channels are marked in colors so as to trace the flow of information.

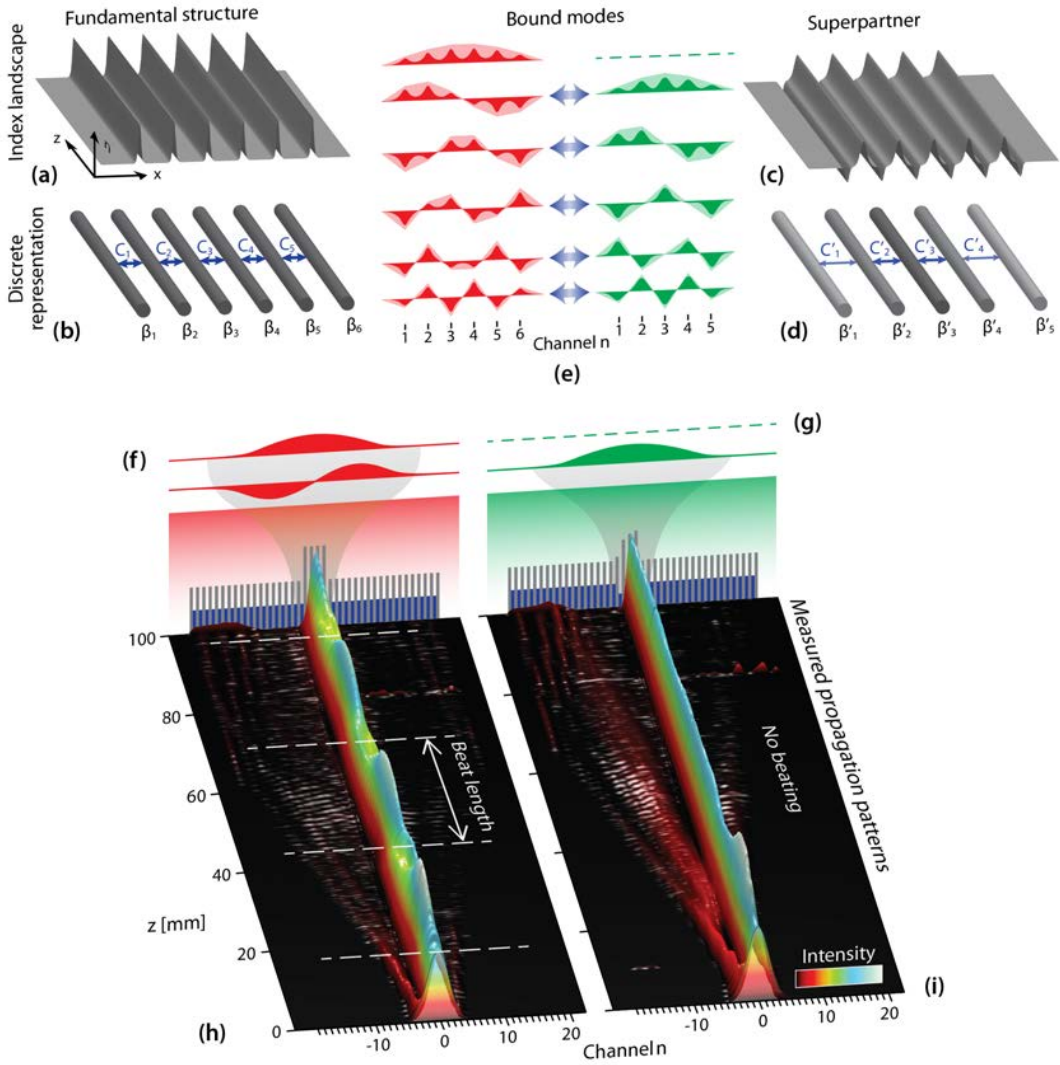

Fig. 2: Supersymmetry in photonic lattices. (a) Continuous refractive index profile and (b) discrete representation of an array arrangement of evenly spaced, identical waveguides and (c,d) its superpartner structure. (e) Bound modes of these two systems. Their vertical position indicates the corresponding eigenvalues; the discrete modes are shown as shaded envelopes. Light dynamics in SUSY structures: The supersymmetric transformation of a two-moded structure (f) yields a single-mode partner (g). Experimentally observed propagation of a Gaussian beam normally injected into a defect domain within a uniform photonic lattice. (h) The emerging periodic beating pattern indicates that the structure supports two bound modes. (i) Propagation of the same initial wave packet in the corresponding superpartner potential. Here, only a single bound state exists and the evolution becomes stationary as soon as the diffractive background has dissipated. In (h,i), the distributions of the propagation constants $\beta$ and coupling coefficients $C$ across the lattices are indicated by gray and blue bars, respectively.

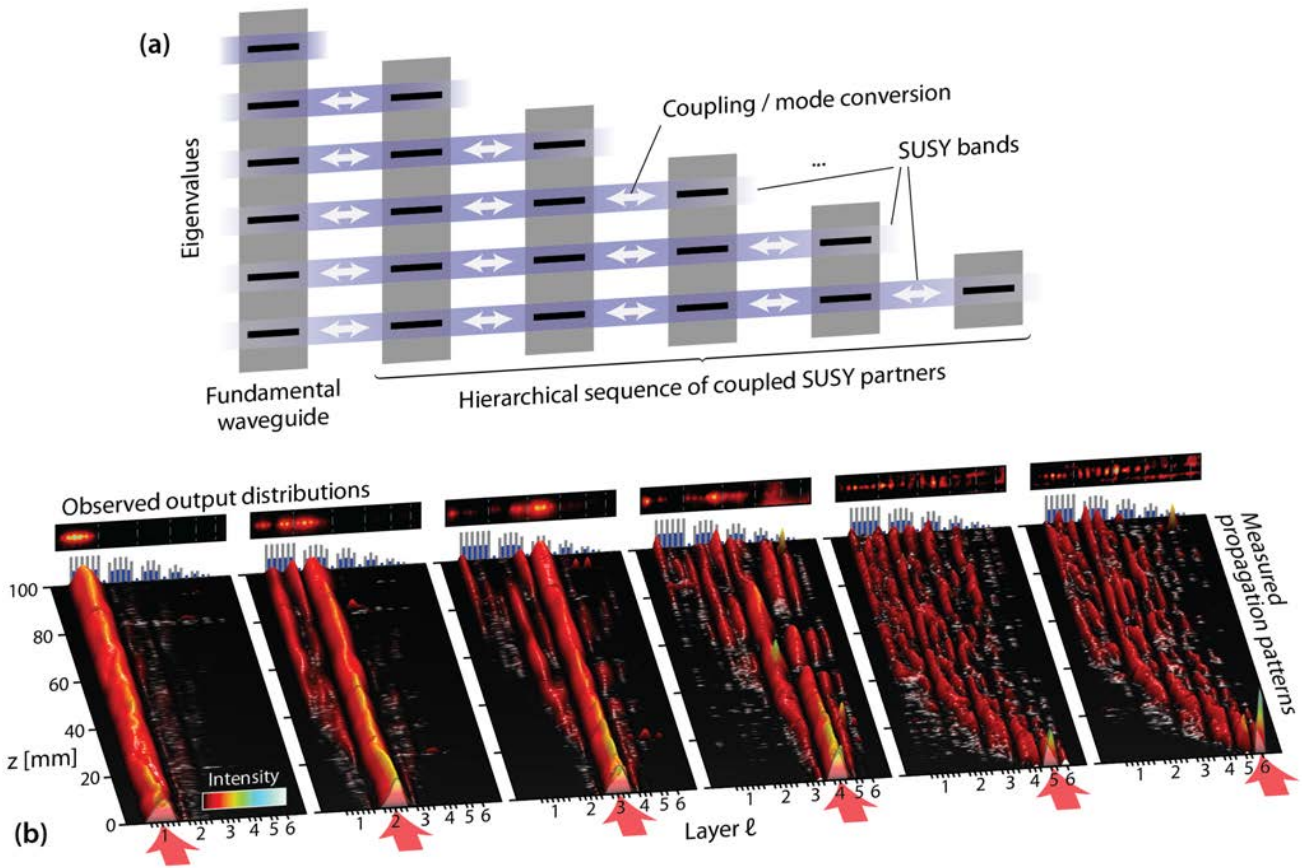

Fig. 3: Supersymmetric Ladder. (a) Schematic representation. A hierarchical sequence of superpartners allows for a stepwise isolation of the higher-order modes of the fundamental system. Coupling between the individual structures allows for a coherent superposition of signals from various inputs. Light traversing the ladder automatically undergoes transformations between the respective modes it occupies. (b) Experimental realization of an optical SUSY ladder with $N_0 = 6$ layers based on photonic lattices. The inter-layer coupling was chosen to be 5% of the coupling in the fundamental array ($C_0 = 0.65\ cm^{-1}$). Shown are plots of the evolving modal intensities obtained by fluorescence microscopy, and images of the intensity distribution at the sample end face. By exciting the fundamental modes of the various layers $\ell_0 = 1 \ldots 6$ (left to right), light is forced to exclusively couple towards lower layers $\ell < \ell_0$.

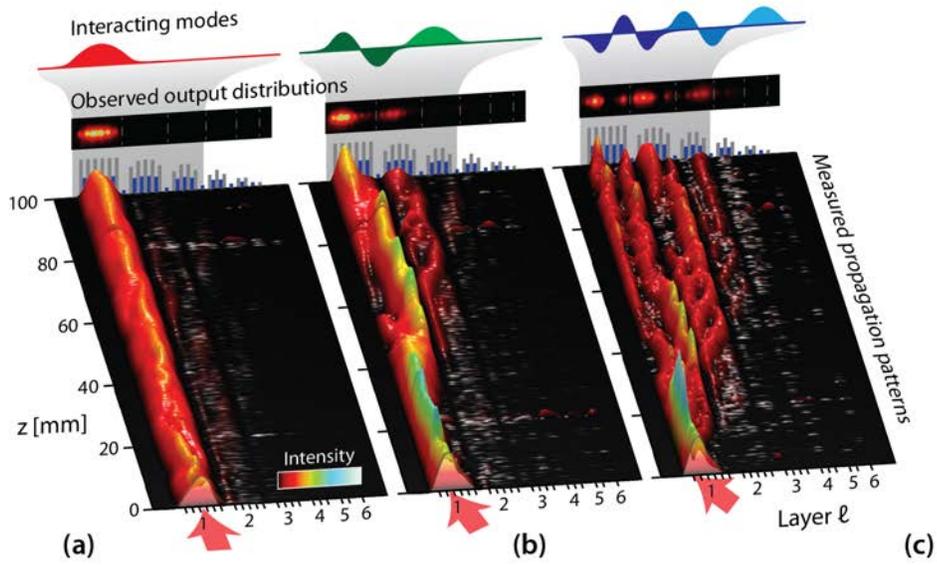

Fig. 4: Mode separation in a SUSY ladder. (a) Light injected into the ground mode of the fundamental layer remains trapped. When a superposition of modes is excited, their respective contents couple to the subsequent layers, as observed for a mixture of the (b) two and (c) three lowest modes. Note that in all cases, the highest accessible layer is exclusively populated in its node-free ground mode. The mixtures of modes in (b) and (c) were prepared by tilting the input wave front at appropriate angles.

**Acknowledgements**

The authors gratefully acknowledge financial support from NSF (grant ECCS-1128520), AFOSR (grants FA9550-12-1-0148 and FA9550-14-1-0037), the German Ministry of Education and Research (Center for Innovation Competence program, grant 03Z1HN31), Thuringian Ministry for Education, Science and Culture (Research group Spacetime, grant no. 11027-514) and the German-Israeli Foundation for Scientific Research and Development (grant 1157-127.14/2011). M.H. was supported by the German National Academy of Sciences Leopoldina (grant LPDS 2012-01).

**Author Contributions**

M.-A.M., R.E.-G. and D.N.C. conceived the idea of optical SUSY. M.-A.M. and M.H. developed the theory of SUSY mode converters and devised the experiments presented here. S.S. fabricated the samples, M.H. and S.S. performed the measurements and M.H. analyzed the data. D.N.C. and A.S. supervised the project; M.H., M.-A.M., D.N.C., S.N. and A.S. co-wrote the manuscript.

**Competing financial interests**

The authors declare no competing financial interests.